\documentclass[a4paper,11pt]{article}
\pdfoutput=1 

\usepackage{jheppub}

\usepackage{booktabs}
\usepackage{fixltx2e}

\usepackage{amsmath,amssymb}
\usepackage{mathrsfs}
\usepackage{mathtools}
\usepackage{braket}

\usepackage{hyperref}

\usepackage{xspace}

\usepackage{accents}

\makeatletter
 \let\old@startsection=\@startsection
 \let\oldl@section=\l@section
 \renewcommand{\@startsection}[6]{\old@startsection{#1}{#2}{#3}{#4}{#5}{#6\mathversion{bold}}}
 \renewcommand{\l@section}[2]{\oldl@section{\mathversion{bold}#1}{#2}}
\makeatother

\title{\boldmath Three-point functions in $\mathcal{N}=4$ SYM: the hexagon proposal at three loops}

\author[a]{Burkhard Eden}
\author[b]{and Alessandro Sfondrini}

\preprint{\texttt{HU-EP-15/44, HU-Mathematik 2015-11}}

\affiliation[a]{Institut f\"ur Mathematik \& Institut f\"ur Physik, Humboldt-Universit\"at zu Berlin,\\
Zum gro{\ss}en Windkanal 6, D-12489 Berlin, Germany.}
\affiliation[b]{Institut f\"ur Theoretische Physik, ETH Z\"urich,\\
Wolfgang-Pauli-Str. 27, CH-8093 Z\"urich, Switzerland.}

\emailAdd{eden@math.hu-berlin.de}
\emailAdd{sfondria@itp.phys.ethz.ch}

\abstract{Basso, Komatsu and Vieira recently proposed an all-loop framework for the computation of three-point functions of single-trace operators of $\mathcal{N}=4$ super-Yang-Mills, the ``hexagon program''. This proposal results in several remarkable predictions, including the three-point function of two protected operators with an unprotected one in the $SU(2)$ and $SL(2)$ sectors. Such predictions consist of an ``asymptotic'' part---similar in spirit to the asymptotic Bethe Ansatz of Beisert and Staudacher for two-point functions---as well as additional finite-size ``wrapping'' L\"uscher-like corrections. The focus of this paper is on such wrapping corrections, which we compute at three-loops in the $SL(2)$ sector. The resulting structure constants perfectly match the ones obtained in the literature from four-point correlators of protected operators.}

\begin{document} 
\maketitle
\flushbottom

\section{Introduction}
A remarkable advance in the study of holographic~\cite{'tHooft:1993gx}, or gauge/string correspondence~\cite{Maldacena:1997re,Witten:1998qj,Gubser:1998bc} was  the discovery of \emph{integrability} in the planar limit~\cite{'tHooft:1973jz} of the correspondence: the duality's dynamics is severely constrained by infinitely many hidden symmetries. These powerful symmetries can be used to compute the energies of string states, or equivalently the two-point functions of operators of the dual CFT. This was firstly done for the most supersymmetric case of $\text{AdS}_5/\text{CFT}_4$, and more recently for other less supersymmetric 
dual pairs.%
\footnote{See refs.~\cite{Arutyunov:2009ga,Beisert:2010jr,Klose:2010ki,Sfondrini:2014via,vanTongeren:2013gva} for reviews and a list of references.}
The missing ingredient to fully describe generic (non-protected)  states in the planar limit through integrability is to exploit symmetries to compute \emph{three-point} functions. A great deal of effort has been devoted to this problem in AdS$_5$/CFT$_4$, both from the point of view of string theory and of gauge theory~\cite{Okuyama:2004bd,Alday:2005nd,Roiban:2004va,Costa:2010rz,Zarembo:2010rr,Escobedo:2010xs,Escobedo:2011xw,Gromov:2011jh,Janik:2011bd,Kazama:2011cp,Gromov:2012uv, Kostov:2012jr, Kostov:2012yq,Kostov:2012wv,Foda:2013nua,Kazama:2012is,Kazama:2013qsa,Klose:2012ju,Jiang:2014mja,Bajnok:2014sza,Kazama:2014sxa,Jiang:2014cya,Bajnok:2015hla,Basso:2015zoa, Hollo:2015cda,Balitsky:2015tca,Jiang:2015lda,Kristjansen:2015gpa,Arnaudov:2015dea}.

Recently, a crucial development was the proposal from Basso, Komatsu and Vieira (BKV) of an all-loop framework for the computation of three-point functions using integrability: the \emph{hexagon program}~\cite{Basso:2015zoa}. This framework automatically incorporates the weakly-coupled ``tailoring'' procedure~\cite{Escobedo:2010xs,Escobedo:2011xw,Gromov:2011jh,Gromov:2012uv}, but it is ``all-loop'' in nature. It is in fact a generalisation of the integrable bootstrap to three-point functions, with a new fundamental object---the hexagon amplitude---playing the role of the scattering matrix. This can be quite easily used to construct the \emph{asymptotic} part of the structure constant, in the sense of the asymptotic Bethe Ansatz of Beisert and Staudacher for two-point functions~\cite{Beisert:2005fw}. For short operators, corrections due to wrapping effects similar to those familiar from the spectral problem~\cite{Ambjorn:2005wa} should be added. BKV propose an explicit recipe for doing so in a manner reminiscent of L\"uscher corrections~\cite{Luscher:1985dn,Luscher:1986pf}.

In ref.~\cite{Basso:2015zoa} BKV put their proposal to several tests, including the direct comparison of certain structure constants against known weak-coupling ~\cite{Eden:2011we,Eden:2012tu} and strong-coupling \cite{Kazama:2014sxa} results. At weak-coupling, BKV explicitly compute the two-loop, three-point function of one non-BPS operator with two BPS ones, in the $SU(2)$ and $SL(2)$ sectors. This is matched to the field theory results, which have been independently obtained from computing the four-point correlators of BPS operators~\cite{Arutyunov:2001mh,Arutyunov:2002fh,Arutyunov:2003ad,Arutyunov:2003ae,Dolan:2004iy,Eden:2011we,Eden:2012tu,Eden:2012rr}. Since the operators appearing in the three-point function are very short, this two-loop calculation already probes the first ``wrapping'' correction to the BKV asymptotic formula---the ones coming from the edge of the hexagon \emph{opposite} to the non-BPS state.

The aim of this paper is to apply the hexagon approach to compute \emph{three-loop} three-point functions in the $SL(2)$ sector. This calculation will for the first time simultaneously probe the wrapping corrections on all the hexagon edges that have a ``mirror'' kinematics. What is more, a prediction for the structure constant of these operator is known in the literature~\cite{Eden:2012tu}. This yields a further quite non-trivial (and successful!) check of the hexagon proposal.

This paper is structured as follows. In Section~\ref{sec:hexagon} we briefly summarise the BKV proposal, working out explicitly some formulae which we will need later---namely, the ones for the wrapping corrections in the two ``adjacent'' channels. In Section~\ref{sec:3ptfun} we specialise these formulae to the aforementioned three-loop computation and discuss how to evaluate them. In Section~\ref{sec:results} we present our results and conclusions. We relegate the technical details concerning the evaluation of the wrapping corrections to the appendices.

\paragraph{Note added:} shortly after the submission of this pre-print we became aware of an upcoming work by Basso, Goncalves, Komatsu and Vieira where, among other things, these three-loop structure constants are computed using the hexagon approach and successfully matched to gauge theory~\cite{Basso:2015eqa}.

\section{The hexagon proposal}
\label{sec:hexagon}
Let us briefly review the BKV proposal~\cite{Basso:2015zoa}. One starts by cutting a three-string interaction (which has the topology of a pair of pants at leading order) into two pieces ``along the pants' seams''. This results into two patches, each having six distinguished edges---three corresponding to the cuts, and three to half of a ``cuff'' each. These hexagons are the central objects of the proposal.

The hexagon can be decorated with six sets of particles. Along the edges corresponding to ``cuffs'', we will have closed-string excitations. Along the edges which will be glued back to yield the three-point functions, we will have excitations in the ``mirror'' kinematics, similarly to what happens when computing finite-size corrections to the energy of string states. 
Given a three-point function by specifying the physical closed-string excitations at each of its cuffs, we can compute the relative structure constant by summing hexagon amplitudes over all possible ways of distributing the physical excitations over the two pieces of the cuffs, as well as summing over all possible mirror states and integrating over the mirror rapidities.

Cleverly using the $SU(2|2)$ super-symmetry of the hexagon, as well the crossing transformation and imposing scattering factorisation 	\textit{\`a la} Zamolodchikov~\cite{Zamolodchikov:1978xm}, the hexagon amplitude $\mathfrak{h}$ was fixed \emph{exactly} in~\cite{Basso:2015zoa}, at least up to a scalar factor $h(x,y)$. This is constrained by crossing symmetry to satisfy
\begin{equation}
\label{eq:crossing}
h(1/x,y)\,h(x,y)=c(x,y)\,,\qquad
c(x,y)=\frac{x^- - y^-}{x^- - y^+}\frac{1-1/x^+y^-}{1-1/x^+y^+},
\end{equation}
where we crossed the Zhukovski variables as $x^\pm\to 1/x^\pm$, corresponding to crossing the rapidity $u\to u^{2\gamma}$~\cite{Basso:2015zoa}. The last ingredient of the BKV proposal is then to set
\begin{equation}
h(x,y)=\frac{x^- - y^-}{x^- - y^+}\frac{1-1/x^-y^+}{1-1/x^+y^+}\frac{1}{\sigma(x,y)}\,,
\end{equation}
where $\sigma$ is the dressing factor of Beisert, Eden and Staudacher~\cite{Beisert:2006ez}.

\subsection{Asymptotic three-point function}
\label{sec:hexagon:asympt}
Using these ingredients, BKV predict the asymptotic part of the three-point function of two protected and one non-protected operator in the $SL(2)$ sector to be
\begin{equation}
\left(\frac{C^{\bullet\circ\circ}_{123}}{C^{\circ\circ\circ}_{123}}\right)^2=\frac{\prod_{k=1}^S \mu(u_k)}{\text{det}\partial_{u_j}\phi_k \prod_{j<k}S_{jk}}
\Big(\sum_{\alpha\cup\bar{\alpha}=\{u\}}\mathcal{A}^{(\alpha,\bar{\alpha})}\Big)^2.
\end{equation}
Here $S_{jk}$ is the $SL(2)$ diagonal  scattering element of the S~matrix by Beisert~\cite{Beisert:2005tm}, $\mu$ is a measure defined by the residue of the pole in the transition $\mathfrak{h}_{D|D}$ of an excitation from one physical edge to another~\cite{Basso:2015zoa},
\begin{equation}
\mu(u) \, = \, i\,\Big(\text{res}_{\substack{v=u}}\, \mathfrak{h}_{D|D}(u|v)\Big)^{-1},
\end{equation}
and the determinant is the Gaudin norm defined in terms of $\phi_j$ which satisfies
\begin{equation}
e^{i\phi_j}=e^{ip_jL}\prod_{k\neq j}S_{jk}\,.
\end{equation}
Finally, we should sum over the partitions $\alpha$ and $\bar{\alpha}$ the expression
\begin{equation}
\mathcal{A}^{(\alpha,\bar{\alpha})}=(-1)^{|\bar{\alpha}|}  \prod_{\substack{j<k\\ j,k\in\alpha\cup\bar{\alpha}}} \!\!h_{jk} \  \prod_{k \in\bar{\alpha}} e^{ip_k\ell}\  \prod_{\substack{j\in\alpha\\k \in\bar{\alpha}}} \frac{1}{h_{jk}}\,.
\end{equation}
Here $\ell$ is the separation between the non-protected operator and the others. Denoting the length of the non-protected operator as $L=L_1$, we have $\ell=\ell_{12}=\ell_{31}$ with $\ell_{ij}=\tfrac{1}{2}(L_i+L_j-L_k)$, all indices being distinct.

\subsection{Wrapping effects}
So far we have not accounted for the presence of mirror particles on the edges of the hexagon to be glued. This can be done in a L\"uscher-like approach, where the leading finite-volume contribution is given by allowing at most \emph{a single} mirror particle per edge. Then one has to correct the asymptotic  expression by
\begin{equation}
\mathcal{A}^{(\alpha,\bar{\alpha})}\to
\mathcal{A}^{(\alpha,\bar{\alpha})}+
\delta \mathcal{A}^{(\alpha,\bar{\alpha})}_{12}+
\delta \mathcal{A}^{(\alpha,\bar{\alpha})}_{23}+
\delta \mathcal{A}^{(\alpha,\bar{\alpha})}_{31}.
\end{equation}
Each of the $\delta\mathcal{A}$ is related to one of the mirror channels. They are given by
\begin{equation}
\label{eq:wrapping}
\delta \mathcal{A}^{(\alpha,\bar{\alpha})}_{jk}=
\sum_{a>0}\int\frac{\text{d}u}{2\pi}
\mu_a^\gamma(u)\,\Big(\frac{1}{x^{[+a]}x^{[-a]}}\Big)^{\ell_{jk}}
\text{int}_a^{(2j-1)\gamma}(u|\{u_i\}).
\end{equation}
Here and in what follows $a$ denotes the bound-state number, and $x^{[\pm a]}$ are the bound-state Zhukovski variables, which depend on the shifted rapidities~$u\pm\frac{a}{2}i$. The mirror measure is the same in all channels and reads
\begin{equation}
\mu_a^\gamma(u)=\frac{a\,(x^{[+a]}x^{[-a]})^2}{g^2(x^{[+a]}x^{[-a]}-1)^2(x^{[+a]}{}^2-1)(x^{[-a]}{}^2-1)}.
\end{equation}
The integrand depends on which mirror channel we consider. Schematically
\begin{equation}
\begin{aligned}
\text{int}^{n\gamma}_a(u,\{u_i\})=&(-1)^{|\bar{\alpha}|}\prod_{j\in\bar{\alpha}}e^{ip_j\ell}\prod_{\substack{i>j\\j\in\bar{\alpha},k\in\alpha}}S_{jk}\\
&\qquad\qquad\qquad\qquad
\sum_{\mathcal{X}_a}(-1)^{f_{\mathcal{X}_a}}\mathfrak{h}_{\mathcal{X}_aD\dots D}(u^{n\gamma},\alpha)\mathfrak{h}_{D\dots D\bar{\mathcal{X}}_a}(\bar{\alpha},u^{-n\gamma}),
\end{aligned}
\end{equation}
which involves the scattering of the mirror (bound-state) particles $\mathcal{X}_a$ with all the physical particles in the $\alpha$ partitions, and similarly for their conjugates $\bar{\mathcal{X}}_a$ with the $\bar{\alpha}$ partition.
 The channel opposite to the non-protected state, corresponding to a shift of $3\gamma$, has been computed in ref.~\cite{Basso:2015zoa}
\begin{equation}
\label{eq:int3gamma}
\text{int}^{3\gamma}_a(u,\{u_i\})=\mathcal{A}^{\alpha,\bar{\alpha}}\frac{(-1)^aT_{a}(u^{\gamma})}{\prod_{j\in\alpha\cup\bar{\alpha}}h_{a}(u^\gamma,u_j)},
\end{equation}
where $T_a$ is the transfer matrix in the anti-symmetric representation (see appendix H in ref.~\cite{Basso:2015zoa}), and $h_a$ is bound-state scalar factor, which can be found by fusion~\cite{Chen:2006gq,Roiban:2006gs,Arutyunov:2009kf}.
The contributions of the two adjacent channels can be easily found and are given by
\begin{equation}
\label{eq:int1gamma}
\text{int}^{\gamma}_a(u,\{u_i\})=\mathcal{A}^{\alpha,\bar{\alpha}}(-1)^aT_{a}(u^{-\gamma}) \prod_{j\in\alpha}\frac{h_{a}(u^\gamma,u_j)}{c_a(u^{-\gamma},u_j)}\prod_{j\in\bar{\alpha}}\frac{h_{a}(u^\gamma,u_j)}{c_a(u^{+\gamma},u_j)},
\end{equation}
and
\begin{equation}
\label{eq:int5gamma}
\text{int}^{5\gamma}_a(u,\{u_i\})=\mathcal{A}^{\alpha,\bar{\alpha}}(-1)^aT_{a}(u^{-\gamma}) \prod_{j\in\alpha}\frac{h_{a}(u^\gamma,u_j)}{c_a(u^{+\gamma},u_j)}\prod_{j\in\bar{\alpha}}\frac{h_{a}(u^\gamma,u_j)}{c_a(u^{-\gamma},u_j)},
\end{equation}
where $c_a(u,v)$ can be found from eq.~\eqref{eq:crossing} by fusion.

There are two important differences between eq.~\eqref{eq:int3gamma} and (\ref{eq:int1gamma}--\ref{eq:int5gamma}). Firstly, the contributions of the adjacent channels are sub-leading since $T_a(u^\gamma)=O(1)$ while $T_a(u^{-\gamma})=O(g^2)$; for this reason, (\ref{eq:int1gamma}--\ref{eq:int5gamma}) did not contribute in the evaluation of the two-loop $SL(2)$ three-point functions in ref.~\cite{Basso:2015zoa}. Secondly, in (\ref{eq:int1gamma}--\ref{eq:int5gamma}) the sum over partitions and the integration over $u$ do not factor, making the evaluation of these contributions somewhat more involved.

It is worth noting that at higher-loop level there are additional wrapping effects contributing, namely the ones described by ordinary L\"uscher corrections for each of the single-trace operators in the three-point function. Of course in this set-up such corrections only appear for the non-protected operator. In the $SL(2)$ sector, as it is well known~\cite{Kotikov:2007cy,Bajnok:2008bm,Bajnok:2008qj}, such wrapping corrections first appear at four loops, and therefore go beyond the scope of this work.

\section{Three-point functions at three loops}
\label{sec:3ptfun}
Using these ingredients, we can now compute three-loop three-point functions involving twist-two, spin-$s$ operators. Specifically, we pick one twist-two (non-protected) operator $\mathcal{O}_1=\text{tr}(D^sZ^2)$, and two protected operators $\mathcal{O}_2=\text{tr}(\bar{Z}Y)$ and $\mathcal{O}_3=\text{tr}(\bar{Z}\bar{Y})$. This will probe the wrapping contributions to the adjacent channels at $\gamma, 5\gamma$, as well as the next-to-leading-order contribution of the opposite ($3\gamma$) channel. Of course the final result will also depend on the three-loop expansion of the asymptotic term described in section~\ref{sec:hexagon:asympt}, as well as on the loop corrections to the rapidities from the Bethe Ansatz. Accounting for these corrections is straightforward, and we will therefore focus our attention on the wrapping effects.

\subsection{Wrapping in the opposite channel}
The computation of the opposite ($3\gamma$) channel follows the one of ref.~\cite{Basso:2015zoa}, but must include next-to-leading order corrections. Firstly, it is useful to massage a bit the transfer matrix $T_a(u^\gamma)$ from ref.~\cite{Basso:2015zoa}---see appendix~\ref{app:transfer} for details. Then, we find it convenient to  strip a denominator out of $T_a(u^\gamma)$, introducing
\begin{equation}
\label{eq:Ttilde}
\widetilde{T}_{a}(u^\gamma)=\frac{(-1)^a\,T_a(u^\gamma)}{\prod_{j=1}^s \text{den}_a(u^\gamma,u_j)}\,,
\quad
\text{den}_a(u^\gamma,u_j)= (x^{[-a]}-x_j^+)(1-1/x^{[-a]}x_j^-).
\end{equation}
The reason for doing so is that, when plugging $T_a$ into~\eqref{eq:int3gamma}, this denominator will simplify against the rational part of the mirror-string dressing factor as we will  see.
We need to expand $\widetilde{T}_a$ up to $O(g^2)$. On top of the tree-level term computed in~\cite{Basso:2015zoa}, we have a correction coming from the one-loop rapidities and one coming from the explicit $g$-dependence of $\widetilde{T}_a$,
\begin{equation}
\label{eq:Tgamma}
\widetilde{T}_a(u^\gamma) =  \Big(1 + g^2 \sum_{j=1}^s u_i^{(1)} \partial_{u_i}\Big) \widetilde{T}_a^{(0)}(u^\gamma)+ g^2 \mathcal{E}^{(1)}\,\widetilde{T}_a^{(1)}(u^\gamma)+O(g^4).
\end{equation}
In terms of the Baxter polynomials $Q(u)=\prod_{j=1}^s (u-u_j)$ we then have\footnote{The Zhukovski variables $x^\pm$ are particularly convenient when discussing crossing transformations, but for our perturbative computations it is convenient to expand them as customary
$x^\pm \rightarrow \sqrt{2}/g \, u^\pm + \ldots \ $.}  
\begin{equation}
\label{eq:g2Tgamma}
\begin{aligned}
\widetilde{T}_a^{(0)}(u^\gamma)&=Q(u^{[a+1]}) + Q(u^{[-a-1]})-Q(u^{[a-1]}) - Q(u^{[-a+1]}),\\
\widetilde{T}_a^{(1)}(u^\gamma)& = \frac{Q(u^{[a+1]})}{u^{[-a]}} - \frac{Q(u^{[a+1]})}{u^{[+a]}} + \sum_{k=1}^{a-1} \frac{Q(u^{[2k-1-a]}) - Q(u^{[2k+1-a]})}{u^{[2 k-a]}}.
\end{aligned}
\end{equation}
Note that the one-loop transfer matrix is multiplied by the one-loop energy
\begin{equation}
\mathcal{E}^{(1)}=\sum_{j=1}^s\frac{-i}{v_j^2+1/4}=-2iS_1(s)\,,
\end{equation}
where the last expression is the well-known representation of the one-loop, spin-$s$ energy as a harmonic sum~\cite{Kotikov:2007cy}.
In~\eqref{eq:g2Tgamma} one might worry that $\widetilde{T}_a^{(1)}$ seems to have a pole on the real-line when $2k=a$. However, it is easy to see that the summand is regular at $u=0$ when one imposes the zero-momentum or level-matching condition $Q(+i/2)=Q(-i/2)$.

As we mentioned, the denominator $\text{den}_a$  simplifies drastically against the rational part of  the dressing factor $h_a$. In fact, for the purpose of our calculation we can write
\begin{equation}
\label{eq:dressingcorrections}
\begin{aligned}
&\text{den}_a(u^\gamma,v)h_{a}(u^\gamma,v)\approx
\frac{1}{x^+_v}\Big(1+ig^2\mathcal{E}^{(1)}\Psi(u)\Big),\\
&\Psi_a(u)=4\gamma+\psi\big(1+iu^{[-a]}\big)
+\psi\big(1-i u^{[-a]}\big)+\psi\big(1+i u^{[+a]}\big)
+\psi\big(1-i u^{[+a]}\big),
\end{aligned}
\end{equation}
where the digamma functions~$\psi$ come from the expansion of the mirror-string dressing factor~\cite{Arutyunov:2009kf,Bajnok:2009vm}.

The last ingredient for the evaluation of the opposite channel contribution is the product of the mirror measure $\mu(u^{\gamma})$ and of the mirror energy $1/x^{[+a]}x^{[-a]}$ from eq.~\eqref{eq:wrapping}. Since in this case $\ell=1$, we find that this is
\begin{equation}
\label{eq:expansionmu}
\frac{g^4 \, a}{(u^{[+a]} u^{[-a]})^3} \left( 1 + g^2 \left(\frac{2}{(u^{[+a]})^2}  + \frac{1}{u^{[+a]} u^{[-a]}} + \frac{2}{(u^{[-a]})^2} \right) + \ldots \right).
\end{equation}

In order to evaluate this expression we have to perform the integration over $u$, sum over the bound-state number $a$ and an additional sum from eq.~\eqref{eq:g2Tgamma}. To this end, it is necessary to massage the expressions we have found, in a way reminiscent of~\cite{Basso:2015zoa}. We comment on the necessary manipulations in appendix~\ref{app:opposite}.

Eventually, we obtain an explicit expression for the integrand to be integrated over the real line, as well as some residues due to manipulations of the integration contours. However, just as it was the case in~\cite{Basso:2015zoa}, the integral cannot be evaluated analytically---to our knowledge. Numeric integration is however possible to good accuracy, even if some care is necessary in the estimate of the numerical errors.%
\footnote{In particular, we have encountered some issues in this sense when using the \texttt{NIntegrate} algorithms of Wolfram Mathematica.}
Nonetheless we could always reduce the result to rational numbers and $\zeta_3, \zeta_5$ by the \emph{pslq} algorithm \cite{PSLQ}. It is interesting to note that the residues subtract all terms involving even $\zeta$ values, while sometimes not contributing to the pure $\zeta_3$ or $\zeta_5$. Remarkably, the integral arising from corrections to mirror dressing phase (see appendix~\ref{app:opposite:mirror}) is apparently always purely rational once the residues are extracted.

\subsection{Wrapping in the adjacent channels}
Let us proceed as in the previous subsection. We firstly define
\begin{equation}
\label{eq:Tminustilde}
\widetilde{T}_{a}(u^{-\gamma})=\frac{(-1)^a\,T_a(u^{-\gamma})}{\prod_{j=1}^s \text{den}_a(u^{-\gamma},u_j)}\,,
\quad
\text{den}_a(u^{-\gamma},u_j)= (x^{[-a]}-x_j^-)(1-1/x^{[-a]}x_j^+).
\end{equation}
Expanding $\widetilde{T}_{a}(u^{-\gamma})$, we find that the first non-vanishing contribution appears at order $g^2$:
\begin{equation}
\label{eq:g2TminusGamma}
\widetilde{T}_a(u^{-\gamma})=g^2
\mathcal{E}^{(1)}
\Big[ -\frac{Q(u^{[-a+1]})}{u^{[-a]}} + \frac{Q(u^{[a+1]})}{u^{[+a]}}
 + \sum_{k=1}^a \frac{Q(u^{[2k-1-a]}) - Q(u^{[2k+1-a]})}{u^{[2 k-a]}} \Big] .
\end{equation}
As we have discussed, the remaining expressions in eqs.~(\ref{eq:int1gamma}--\ref{eq:int5gamma}) contain terms which are both $u$-dependent and partition dependent. Let us focus on one channel, say $1\gamma$~\eqref{eq:int1gamma}, and introduce the incomplete Baxter polynomial $Q_\alpha(u)=\prod_{j\in\alpha}(u-u_j)$. Using that $Q(u)=Q_{\alpha}(u)Q_{\bar{\alpha}}(u)$ we can \textit{e.g.} eliminate the products over the $\bar{\alpha}$ partition.
The price is to introduce a product over the $\alpha$ partition
\begin{equation}
M_\alpha \, = \, \left( \prod_{j \in \alpha} e^{- 2 i p_j} \right) \frac{Q_\alpha(u^{[a-1]}) Q_\alpha(u^{[-a-1]})}{Q_\alpha(u^{[a+1]}) Q_\alpha(u^{[-a+1]})},
\end{equation}
which modifies the partition dependent term so that it differs from the asymptotic one $\mathcal{A}=\sum_{\alpha\cup\bar{\alpha}} \mathcal{A}^{\alpha,\bar{\alpha}}$.
Instead we now have
\begin{equation}
\widetilde{\mathcal{A}}=\sum_{\alpha\cup\bar{\alpha}}\mathcal{A}^{\alpha,\bar{\alpha}} M_\alpha,
\end{equation}
which depends on $a$ and~$u$. Clearly the $5\gamma$  channel can be found by exchanging $\alpha\leftrightarrow\bar{\alpha}$. As it turns out, $\widetilde{\mathcal{A}}$ is identical in the two cases.

Given that we are interested in integrating this expression, it is convenient to make all possible poles manifest. To this end, we introduce a function $\widetilde Q(u,a,s)$ that satisfies
\begin{equation}
\label{eq:defQtilde}
\widetilde{\mathcal{A}}\ Q(u^{[a+1]})\,Q(u^{[-a+1]})=\mathcal{A}\  \widetilde{Q}(u,a,s).
\end{equation}
By construction, $\widetilde{Q}(u,a,s)$ is a polynomial in $u$, \textit{i.e.}\ it has no poles. Such a polynomial can be explicitly evaluated for every given spin $s$, see appendix~\ref{app:adjacent}.
 Using the expansion of the remaining terms in~\eqref{eq:int1gamma} we find that the integrand takes the form
\begin{equation}
\label{eq:adjacent}
{\cal A} \, \frac{Q(i/2) \ \widetilde Q(u,a,s)}{Q(u^{[a+1]}) Q(u^{[-a-1]}) Q(u^{[a-1]}) Q(u^{[-a+1]})} \, T_a(u^{-\gamma}). 
\end{equation}
Even if $\widetilde Q(u,a,s)$ in general is not a Baxter polynomial, it has degree~$s$. It follows that the integrand decays fast for large $|u|$, and can be evaluated by residues. We further comment on the related technicalities in appendix~\ref{app:adjacent}.

\section{Results, conclusions and outlook}
\label{sec:results}

Summing up the various contributions described in the last two sections we finally obtain the structure constants of two half-BPS operators of length-2 going into a twist-2 operator:
\begin{table}[h!]
\begin{center}
\begin{tabular}{l|l}
\toprule
$s$ & $\left(\frac{C^{\bullet \circ \circ}}{C^{\circ \circ \circ}}\right)^2$ for twist $L=2$, bridge $\ell=1$ and spin $s$ \\[1 mm]
\midrule
2 & $\frac{1}{6}-g^2+(7+3 \zeta_3) g^4-(48+8 \zeta_3+25 \zeta_5) g^6+ \ldots$ \\[2 mm]
4 &  $\frac{1}{70}-\frac{205}{1764}g^2+\left(\frac{76393}{74088}+\frac{5}{14} \zeta_3 \right)
   g^4
   - \left(\frac{242613655}{28005264} + \frac{1315}{1323} \zeta_3+ \frac{125}{42}\zeta_5 \right) g^6
   + \ldots$ \\[2 mm]
6 & $\frac{1}{924}-\frac{553}{54450} g^2 +\left(\frac{880821373}{8624880000}+\frac{7}{220} \zeta_3 \right)
   g^4-\left(\frac{1364275757197}{1423105200000}+\frac{520093}{6534000} \zeta_3 +\frac{35}{132} \zeta_5 \right)
   g^6+ \ldots$ \\[2 mm]
8 & $\frac{1}{12870}-\frac{14380057}{18036018000} g^2 +\left(\frac{5944825782678337}{682443241880400000}+\frac{761}{300300} \zeta_3 \right)
   g^4 $ \\[1 mm]
   & $- \left(\frac{758072803634287465765957}{8607383632540733040000000}
   +\frac{15248925343}{2840672835000}\zeta_3+\frac{761}{36036} \zeta_5 \right)
   g^6+ \ldots$ \\[2 mm]
10 & $\frac{1}{184756}-\frac{3313402433}{55983859495200} g^2+\left(\frac{171050793565932326659}{248804677619932
   936320000}+\frac{671}{3527160} \zeta_3 \right)
   g^4$ \\[2 mm]
   & $-\left(\frac{9135036882706194334305789554347}{1243961012766985364
   412864576000000}+\frac{11482697774339}{35269831481976000} \zeta_3+\frac{3355}{2116296} \zeta_5 \right)
   g^6+ \ldots$
\end{tabular}
\end{center}
\end{table}
The $O(g^6)$ values  in the table are in exact agreement with the conformal partial-wave analysis of the four-point function of stress energy tensor multiplets \cite{Eden:2011we,Eden:2012rr}. This test of the hexagon conjecture \cite{Basso:2015zoa} was the main motivation for our work. It probes the correctness of the new approach to finite size corrections already rather deeply, as is well illustrated by the multitude of effects we had to take into account.

An obvious extension to this work is to predict structure constants for the fusion of two higher-length single-trace half-BPS operators into twist 2 operators. In that case the bridge length for the adjacent channels would stay put at $\ell_{12} = \ell_{31} = 1$, while $\ell_{23}$ for the opposite channel would rise. For length-3 operators we have $\ell_{23} = 2$, so that the leading order analysis of \cite{Basso:2015zoa} now applies to the three-loop correction. As the mirror measure now starts on $1/(u^+ u^-)^4$ we obtain a modified effective integration measure
\begin{eqnarray}
\hat \mu & = & \frac{16 \pi^3}{3 (1 + 
   4 u^2)^4 \cosh^2(\pi u)} \biggl[ \frac{\pi (-1 + 8 u^2 + 
      48 u^4)}{\cosh^2(\pi u)} \\ & & \qquad\qquad\qquad + 48 u (-1 + 4 u^2) \tanh(\pi u) - 
   2 \pi (-1 + 8 u^2 + 48 u^4) \tanh^2(\pi u) \biggr] \nonumber \, .
\end{eqnarray}
with the subtraction of residues following the by now standard path. The contribution of the opposite channel for length 3 is marked in the table by the coefficient $\eta$, which ought to be put to $1/2$ in this case. If the length of the BPS is greater or equal 4, we have bridge length $\ell_{23} > 2$ and the leading contribution in the opposite channel moves out to $O(g^8)$ or higher. The approach of \cite{Basso:2015zoa} then predicts the result in the table below at $\eta=0$, so notably perfect universality of the structure constants up to three loops. We will check what constraints these results can impose on Ans\"atze for higher-charge planar correlation functions~\cite{Eden:2011we}.

\begin{table}[ht]
\begin{center}
\begin{tabular}{l|l}
\toprule
$s$ & $\left(\frac{C^{\bullet \circ \circ}}{C^{\circ \circ \circ}}\right)^2$ for twist $L=2$, bridge $\ell_{12}=\ell_{31}=1$, $\ell_{23} > 1$ and spin $s$ \\[1 mm]
\midrule
2 & $\frac{1}{6}-g^2+7 g^4+(10 \zeta_5 \eta-10 \zeta_5+7 \zeta_3 -48)
   g^6+\ldots$ \\[2 mm]
4 &  $\frac{1}{70}-\frac{205}{1764}g^2+\frac{36653}{37044}g^4+\left( \left(\frac{1}{6} \zeta_3+ \frac{25}{21}\zeta_5\right) \eta-\frac{25}{21}\zeta_5+\frac{193}{216} \zeta_3-\frac{442765625}{56010528}\right)
   g^6+\ldots$ \\[2 mm]
6 & $\frac{1}{924}-\frac{553}{54450}g^2+\frac{826643623}{8624880000}g^4$ \\[2 mm]
& $+\left(\left(-\frac{1}{1440}+\frac{7}{264} \zeta_3+\frac{7}{66} \zeta_5\right) \eta-\frac{7}{66} \zeta_5+\frac{24143}{297000} \zeta_3-\frac{1183056555847}{1423105200000}\right)
   g^6 + \ldots$ \\[2 mm]
8 & $\frac{1}{12870}-\frac{14380057}{18036018000}g^2+\frac{2748342985341731}{341221620940200000}g^4+\bigl(\left(-\frac{79}{604800}+\frac{3}{1040}\zeta_3+\frac{761}{90090} \zeta_5 \right) \eta$ \\[2 mm]
& $-\frac{761}{90090} \zeta_5+\frac{1039202363}{158918760000} \zeta_3-\frac{1270649655622342732745039}{1721476726508146608000000
   0}\bigr)
   g^6+\ldots$ \\[2 mm]
10 & $\frac{1}{184756}-\frac{3313402433}{55983859495200} g^2 +\frac{156422034186391633909}{248804677619932936320000} g^4 +\bigl( \bigl(-\frac{45071}{2813045760} + \frac{781}{2930256} \zeta_3 +$
\\[2 mm] & $\frac{671}{1058148} \zeta_5\bigr) \eta -\frac{671}{1058148} \zeta_5 +\frac{8295615163}{16799157648000} \zeta_3-\frac{7465848687069712820911408164847}{12439610127669853
   64412864576000000}\bigr) g^6+\ldots$
\end{tabular}
\end{center}
\end{table}

It would be very interesting to extend this analysis to higher-order
corrections. A first question is how one can make sense of wrapping
corrections to the operators of the three-point functions. While this is a
familiar problem in the context of two-point functions, this issue still to
be explored in the hexagon program. As we mentioned, these effects would
first appear at order $O(g^8)$, and therefore are accessible to
gauge-theoretical computations, which would provide another crucial check of
the hexagon approach.

In  the long run one could try to constrain a putative octagon operator \cite{Basso:2015zoa} for four-point functions by perturbative data. Ideally we obtain a machinery that will directly furnish non-trivial kinematics, so which will allow us to reach out beyond the computation of sets of constants.

Another difficult but very interesting question is whether the hexagon
approach can be promoted to a truly non-perturbative formalism. While as we
saw L\"uscher-like corrections work remarkably well, one should also account account
for more than one virtual particle at a time%
\footnote{For advances in this direction in the context of
two-point functions see ref.~\cite{Bombardelli:2013yka}.} within the hexagon approach~\cite{Basso:2015zoa}.
Ideally one would hope that a sort of ``Thermodynamic Bethe Ansatz''
formulation could be constructed for the hexagon. We are confident to
witness remarkable developments in this direction in the near future.

\acknowledgments
We would like to thank G.~Arutyunov and S.~van Tongeren for discussions. B.E. is supported by the DFG, ``eigene Stelle" ED 78/4-2 and acknowledges partial support by the Marie  Curie network GATIS under REA Grant Agreement No 317089.
A.S.~would like to thank the group for Mathematical Physics of Space, Time and Matter at Humboldt University where part of this research was carried out. A.S.'s research was partially supported by the NCCR SwissMAP, funded by the Swiss National Science Foundation.

\appendix
\section{Telescoping the transfer matrix}
\label{app:transfer}
The transfer matrix in the antisymmetric bound state representation as defined in formula (H1) in ref.~\cite{Basso:2015zoa} can be substantially simplified when the level-matching condition, or, in gauge theory parlance, the zero momentum condition $\prod_j \, x_j^+ / x_j^- = 1$ is satisfied. Using
\begin{equation}
(x^{[a]} - x_j^\mp ) \left( 1 - \frac{1}{ \, x^{[a]}  x_j^\mp} \right) =  Q(u^{[a \pm 1]})
\end{equation}
it follows that
\begin{equation}
\frac{R^+(u^{[a]}) \, B^+(u^{[a]})}{R^-(u^{[a]}) \, B^-(u^{[a]})} \, = \, \frac{Q(u^{[a+1]})}{Q(u^{[a-1]})},
\end{equation}
where we followed the notation of~\cite{Basso:2015zoa}
\begin{equation}
R^\pm(u) \, = \, \prod_j (x(u) - x_j^\mp) \, , \qquad B^\pm(u) \, = \, \prod_j \left(\frac{1}{x(u)} - x_j^\mp\right) \, .
\end{equation}
In every summand  in $T_a(u^\gamma)$ and $T_a(u^{-\gamma})$ the $Q$ factors ``telescope'', so that all terms but the first in the denominator and the last in the numerator cancel. In terms of the function $\widetilde{T}_a$ defined in~\eqref{eq:Ttilde}
\begin{equation}
\label{eq:defT}
\begin{aligned}
\widetilde{T}_a(u^\gamma)=& R^-(u^{[-a]})B^{+}(u^{[-a]})+R^+(u^{[a]})B^{-}(u^{[a]})-2R^-(u^{[a]})B^{-}(u^{[a]})\\
&+\sum_{k=1}^{a-1}\Big(R^+(u^{[2k-a]})B^-(u^{[2k-a]})+R^-(u^{[2k-a]})B^+(u^{[2k-a]})\\
&\qquad\qquad\qquad\qquad\qquad\qquad\qquad\qquad-2R^-(u^{[2k-a]})B^-(u^{[2k-a]})\Big).
\end{aligned}
\end{equation}
A similar expression for $\widetilde{T}_a(u^\gamma)$ can be found immediately by crossing $x^{[\pm a]}\to 1/x^{[\pm a]}$, \textit{i.e.}\ by swapping $B^\pm\leftrightarrow R^{\pm}$. 

\section{Evaluation of the opposite-channel wrapping}
\label{app:opposite}
With respect to the computation detailed in ref.~\cite{Basso:2015zoa} we now have to include sub-leading order $O(g^6)$ contributions. These may come from corrections to different bits of the leading-order expansion of eq.~\eqref{eq:wrapping}:
\begin{enumerate}
\itemsep0pt
\item From corrections to the rapidities, through $\sum_{j=1}^s u^{(1)}_j\partial_{u_j}$. This does not substantially alter the analytic form of leading-order expression, and can be easily evaluated by shifting the integration variable to reduce the integral to take values over a single $Q$-function~\cite{Basso:2015zoa}.
\item From corrections to the integration measure $\mu(u^\gamma)$, \textit{cf}.\ eq.~\eqref{eq:expansionmu}.
\item From corrections  $\widetilde{T}^{(1)}_a(u^\gamma)$ to the transfer matrix $\widetilde{T}_a(u^\gamma)$, \textit{cf}.\ eq.~\eqref{eq:g2Tgamma}.
\item From corrections coming from the dressing factor, \textit{cf}.\ eq.~\eqref{eq:dressingcorrections}.
\end{enumerate}
Below we will discuss in more detail these last three contributions.

\subsection{Corrections to the measure}
The strategy here is once again to perform shifts in the integration variable~$u$ in such a way as to end up with a a single $Q$-function $Q(u)$. These  shifts produce the total effective measure
\begin{equation}
\begin{aligned}
\mu^c & =  \frac{16 \pi^2}{(3 (1 + 4 u^2)^5)  \, \cosh^2(\pi u)}\,\Bigl[ 24 (1 - 40 u^2 + 80 u^4)  \\
& \qquad + 48 \pi u (1 - 16 u^4) \tanh(\pi u) - 
  4 \pi^3 u (1 + 4 u^2)^3 \tanh^3(\pi u)   \\
   &\qquad+\pi^2 (1 + 
     4 u^2)^2 \cosh^{-2}(\pi u) \bigl[(-1 + 12 u^2) (-2 + \cosh(2 \pi u))\\
 &\qquad\qquad\qquad\qquad\qquad\qquad\qquad\qquad + 
     8 \pi u (1 + 4 u^2) \tanh(\pi u)\bigr] \Bigr].
\end{aligned}
\end{equation}
Like the leading measure this falls off  exponentially for large $|u|$ so that integration against the Baxter polynomials $Q(u)$ is possible for any spin. The shifts $u \rightarrow u \pm \tfrac{a+1}{2}i$ must again be accompanied by the subtraction of residues from crossing the poles at $\mp \tfrac{a}{2}i$.

\subsection{Corrections to the transfer matrix}
Let us consider $\widetilde{T}^{(1))}_a(u^\gamma)$~\eqref{eq:g2Tgamma}. Once again, we want to shift~$u$ in such a way as to integrate only on $Q(u)$.  For the boundary terms with $Q(u^{[-a-1]}),$ $Q(u^{[a+1]})$ this is quite as before. The sum over $k$ can be concisely rewritten: originally one has
\begin{equation}
Q(u) \sum_{a=1}^\infty a \sum_{k=1}^{a-1} \, \frac{1}{u^+ \, \left((a-k) - i  \, u^+ \right)^3 \, \left(k + i \, u^+ \right)^3} - (u^+ \leftrightarrow u^-).
\end{equation}
We can swap the order of summations and rewrite the sum in terms of $a' = a-k$. In this way, the sums decouple and we can evaluate in terms of polygamma functions.

The ``effective'' integration measure emerging form these shifts is 
\begin{equation}
\begin{aligned}
\mu^p  =&
\frac{32 i \pi^2}{(1 + 4 u^2)^4 \, \cosh^2(\pi u)}
\left(1 - 24 u^2 + 16 u^4 + \pi u (3 + 8 u^2 - 16 u^4) \tanh(\pi u) \right)  \\
& \qquad+ \frac{2 i}{1 + 4 u^2} \Big( \psi' \left( i u^- \right) \, \psi'' \left(- i u^+ \right) + \psi' \left(- i u^+ \right) \, \psi'' \left(i u^- \right) \Big).
\end{aligned}
\end{equation}
The terms in the second line cannot straightforwardly be rewritten in terms of trigonometric functions; nonetheless this part also has the desired asymptotic behaviour $e^{- 2 \pi |u|}$ for large values of the argument.

Residues from the boundary terms in the second line of (\ref{eq:g2Tgamma}) are computed as usual.
A further remark concerns the subtraction for the sum part of (\ref{eq:g2Tgamma}): to begin with let $a$ be odd. Then the shift of $Q(u^{-a+2 k - 1}) \rightarrow Q(u)$ crosses the denominator pole only if $-a + 2 k < 0$, likewise for $-a + 2 k > 0$ only the other term picks up a residue. For even $a, \, k = a/2$ we look at
$(Q(u^-) - Q(u^+))/u$. In either term the shift approaches the pole but does not cross it. As in a principle value prescription both terms contribute half a residue. Therefore in both cases---even and odd bound state number---we have to subtract
\begin{equation}
i\,Q\left(i/2\right) \sum_{a=1}^\infty \sum_{k = 1}^{a-1} \frac{a}{(a-k)^3 \, k^3} \, = \, 2 i \, \zeta_2 \zeta_3 \, Q\left(i/2\right),
\end{equation}
where the right-hand-side can be found by the same manipulations on the double sum as above.

\subsection{Measure and residues from the first order mirror dressing phase}
\label{app:opposite:mirror}
Beyond a rational factor the mirror dressing phase contains the very special combination of digamma functions $\Psi_a(u)$ of eq.~\eqref{eq:dressingcorrections}. This results in an integrand of the form
\begin{equation}
S_1(s) \sum_{a>0} \, \frac{a\,\Psi_a(u)}{(u^{[+a]} \, u^{[-a]})^3} \, \big( Q(u^{[a+1]}) + Q(u^{[-a-1]}) - Q(u^{[a-1]}) - Q(u^{[-a+1]}) \big).
\end{equation}
Once again the strategy is to shift each term in the integrand in such a way as to extract an overall factor of~$Q(u)$. The presence of the digamma functions mandates some extra care. For each shift, two of the four digamma functions~$\psi$ in~$\Psi_a$ become independent of the bound state number $a$. The sum over $a$ then simply yields a second polygamma factor. 
As for the digamma functions that do depend on the bound~state number even after shifting, we recall that
\begin{equation}
\label{eq:psidef}
\psi(1 + x) \, = \, \sum_{k=1}^\infty \Big(\frac{1}{k}-\frac{1}{k+x}\Big) - \gamma,
\end{equation}
which leads to a second summation over terms of the type $1/(a-k+y)$ or $1/(a+k+y)$, as well as formally some $\zeta(1)$ terms. One of
the factors $(u^{[+]a})^{-3}$ or $(u^{[-a]})^{-3}$ from the bound state measure yields $1/(a \pm 1/2 \pm i u)^3$, too. The idea is now to use partial fractions w.r.t. $a$ to decompose products of the two factors involving $a$ into single (if higher order) poles. We then swap the sums as before and shift to $a' = a-k$ and $a'' = a + k$. The sums from $1-k, k+1$ to infinity, respectively, are conveniently split into an infinite part with bounds $1 \ldots \infty$ and a finite bit $-a' \in \{0 \ldots k-1\}$ and $a'' \in \{1 \ldots k\}$. After some algebra it is found that all $\zeta(1)$ terms cancel and what is more, the finite sums ``telescope'' whereby the nested double sum totally disappears. The net result of this complicated looking exercise is surprisingly concise
\begin{equation}
\frac{\gamma}{(u^-)^3} \Bigl[ 2 i \psi'(1 - i u^-) + 2 i \psi'(1 + i u^-) + 
 u^- \psi''(1 - i u^-) - u^- \psi''(1 + i u^-) \Bigr] - (u^- \leftrightarrow u^+) \, .
\end{equation}
Adding it to the contribution from the $a$ independent digamma functions we obtain the complete effective measure 
\begin{equation}
\begin{aligned}
\mu^m = & \frac{32 \pi^2}{(1 + 
  4 u^2)^4 \, \cosh^2(\pi u)}
\Bigl[-1 + 24 u^2 - 16 u^4 +\\
&\qquad \qquad \qquad \qquad \qquad \qquad \pi u (-3 - 8 u^2 + 16 u^4) \tanh(\pi u) \Bigr] \\
& - \frac{16 \pi^2}{(1 + 4 u^2)^3 \, \cosh^2(\pi u)} 
\Bigl[1 - 12 u^2 +  2 \pi u (1 + 4 u^2) \tanh(\pi u) \Bigr]\\
&\qquad \qquad \qquad \qquad \qquad \qquad\qquad\times 
\Bigl[2 \gamma + \psi(- i u^+) + \psi( i u^-) \Bigr],
\end{aligned}
\end{equation}
induced by the mirror dressing phase.

Here the subtraction of residues deserves further attention. To begin with note that $\Psi_a$ has simple poles at  $\pm i (a/2 + k)$. Shifting thus needs to be complemented by the subtraction of residues only at $\pm i a/2$.
Let us focus on the $Q(u^{[a+1]})$ polynomial, \textit{i.e.}\ on the residue at $u = + i a/2$. We decompose
\begin{equation}
\Psi_a(u) \, = \, \Psi_a^{\text{reg}}(u) + \frac{i}{u - \frac{i}{2} a}
\end{equation}
by adding and subtracting the pole. Now, the pole part of $\Psi(u)$ will combine with the $(u^{[-a]})^3$ factor from the bound-state mirror measure to form a fourth-order pole. This residue is then given by a third derivative acting on $Q(u^{[a+1]})/(u^{[+a]})^3$.

Secondly, the bound-state measure itself has the familiar third-order pole, with now a residue $\Psi_a^{\text{reg}}(u) \, Q(u^{[a+1]}) / (u^{[a]})^3$. It is not hard to work out that
\begin{equation}
\begin{aligned}
\Psi_a^{\text{reg}}|_{\frac{i}{2} a} = & \ \ 2 S_1(a) - \frac{1}{a} \, ,  \\
\Psi_a^{\text{reg}}{}'|_{\frac{i}{2} a} = & i \left(2 S_2(a) - \frac{1}{a^2} \right) , \\
\Psi_a^{\text{reg}}{}''|_{\frac{i}{2} a} = & - 2 \left(2 S_3(a) - \frac{1}{a^3} - 4 \zeta(3) \right).
\end{aligned}
\end{equation}
The bound state sum therefore creates all the double $\zeta$-values $\{\zeta_{21},$ $\zeta_{22},$ $\zeta_{23},$ $\zeta_{31},$ $\zeta_{32},$ $\zeta_{41}\}$ up to transcendentality weight 5. All of these can be recast in terms of ordinary zeta-values.

\section{Evaluation of the adjacent-channel wrapping}
\label{app:adjacent}
The crucial ingredient in the evaluation of the adjacent-channel wrapping is the polynomial~$\widetilde Q(u,a,s)$ introduced in~\eqref{eq:defQtilde}. This is a degree-$s$ polynomial that can be explicitly evaluated for each given~$s$. Interestingly, if we extract a normalisation pre-factor $\widetilde Q=n \widetilde Q'$ so that the highest-degree monomial in $\widetilde Q'$ has unit coefficient, we find $n=Q(i/2)$. We collect the first few expression for $\widetilde Q$ in this normalisation in the table below. Curiously, when $a = 1$, the $\tilde Q$s are in fact Baxter polynomials.

\begin{table}[h!]
\begin{center}
\begin{tabular}{l|l}
\toprule
spin & $\widetilde Q(u,a,s)/Q(i/2)$ \\[1mm]
\midrule
2 & $-\frac{1}{3} + \frac{1}{4} a^2 + u^2$ \\[1mm]
4 & $ \ \frac{12}{35} - \frac{23}{84} a^2 - \frac{1}{48} a^4 - \frac{23}{21} u^2 + \frac{1}{6} a^2 u^2 + u^4$  \\[1 mm]
6 & $-\frac{60}{77} + \frac{139}{220}a^2 + \frac{3}{44}a^4 + \frac{1}{320}a^6 + \frac{139}{55}u^2 - \frac{
 6}{11}a^2 u^2 - \frac{9}{80}a^4 u^2 - \frac{36}{11}u^4 - \frac{1}{4}a^2 u^4 + u^6$
\end{tabular}
\end{center}
\end{table}%

Let us now see how to integrate~\eqref{eq:adjacent}.
As had been mentioned above, in the sum part of (\ref{eq:g2TminusGamma}) the poles $1/u^{-a+2 k}$ are absent because the difference in the respective numerator factors out a power of $u^{-a+ 2 k}$. Since we are not interested in shifting $u$ it is best to cancel these factors and to work with the remaining total numerator polynomials. In these it is of course trivial to execute the sum over $k$. Note that $\widetilde Q$ times this polynomial or $Q(u^{[a-1]}), \, Q(u^{[-a+1]})$ contains powers up to $a^{2 s}$, to be augmented by the explicit factor $a$ in the bound state measure.

Any residue has various denominator terms $(a + x)^n$ where $x$ may contain $\pm 1/2, \pm 1$ and/or one or two Bethe roots. We may now use partial fractions w.r.t. to $a$ to reduce to single (generically higher order) poles which can be summed over the bound state counter to yield polygamma functions. The procedure is well-behaved in that 
\begin{enumerate}
\itemsep0pt
\item all potentially divergent sums $\zeta_0, \zeta_{-1} \ldots \zeta_{-2 s-1}$ cancel,
\item $\zeta_1$ also cancels when the sums $1/(a+x)$ are expressed as $\psi(1+x)$ by eq. (\ref{eq:psidef}),
\item  and the transcendentality level does not increase with the spin.
\end{enumerate}
One then finds some $\zeta$-values with rational coefficients and a sum over polygamma functions containing the Bethe roots in their arguments. For spin 2 this can immediately be simplified to rational numbers and $\zeta_3, \zeta_5$ due to the property $\psi(1+x) = \psi(x) + 1/x$ and its derivatives. In general, we have not tried to analytically solve this---which may well be possible---but rather numerically evaluated to high precision and reconverted by the \emph{pslq}  algorithm. The adjacent channel calculation has a definite advantage on the opposite channel case because no integration is needed to arrive at the final result whereby the numerical precision can be much better.

\bibliography{bib-3pt}{}
\bibliographystyle{JHEP}

\end{document}